\documentclass[journal=jpccck,manuscript=article,layout=twocolumn]{achemso}
\pdfoutput=1
\usepackage[utf8]{inputenc}

\usepackage{chemformula} % Formula subscripts using \ch{}
\usepackage{amsmath,amssymb,amsfonts,mathtools,physics}
\usepackage{dblfloatfix}

\newcommand{\kB}{k_{\rm\scriptscriptstyle B}}


\author{Marco Bosi}
\affiliation{ 
Universit\"at Osnabr{\" u}ck, 
Fachbereich Physik, 
Barbarastra{\ss}e 7, 
D-49076 Osnabr\"uck, 
Germany}

\author{Julian Fischer}
\affiliation{ 
Universit\"at Osnabr{\" u}ck, 
Fachbereich Physik, 
Barbarastra{\ss}e 7, 
D-49076 Osnabr\"uck, 
Germany}

\author{Philipp Maass}
\email{maass@uni-osnabrueck.de}
\affiliation{ 
Universit\"at Osnabr{\" u}ck, 
Fachbereich Physik, 
Barbarastra{\ss}e 7, 
D-49076 Osnabr\"uck, 
Germany}

\title{Network-Forming Units, Energy Landscapes, and Conductivity Activation Energies in Alkali Borophosphate Glasses: Analytical Approaches}

\SectionNumbersOn 
\begin{document}

%\begin{tocentry}
%\centering
%%.Size: 3.25 in. × 1.75 in. (approx. 8.25 cm × 4.45 cm). 
%\includegraphics[scale=1]{tocgraphic}\\[1.5cm]  
%\end{tocentry}

\begin{abstract}
A major challenge in the modeling of ionically conducting glasses is to
understand how the large variety of possible chemical compositions and
specific features of their structure influence ionic transport
quantities. Here we revisit and extend a theoretical approach for
alkali borophosphate glasses, where changes of conductivity activation
energies with the borate to phosphate mixing ratio are related to
modifications of the ionic site energy landscape. The landscape
modifications are caused by varying amounts of different units forming
the glassy network, which lead to spatial redistributions of the
counter-charges of the mobile alkali ions. Theoretical
approaches are presented to calculate variations of both network
former unit concentrations and activation energies with the glass composition. 
Applications to several alkali borophosphate glasses show good agreement with
experimental data.
\end{abstract}


\section{Introduction}
\label{sec:introduction}
High ionic conductivities are desired when optimizing glassy
electrolytes for most applications \cite{Ingram:1987}. To guide this
optimization process, it is important to improve our theoretical
understanding of ion motion in glasses \cite{Dyre/etal:2009}.
Strong enhancements of ionic conductivities can be achieved by
utilizing various effects, which we term the ``ionic concentration effect'',
``halide doping effect'', and ``mixed glass former effect''. For the
ionic concentration effect, a strong (superlinear) increase of ionic conductivity results from
an increase of the molar content of mobile ions,\cite{Hakim/Uhlmann:1971} and for the 
halide doping effect from
an addition of alkali halides  \cite{Kawamura/Shimoji:1989}. 
In the case of the ``mixed glass former effect'', \cite{Desphande/etal:1988,
  Jayasinghe/etal:1996-p} a mixing of different types of glass
formers, such as silicates, borates, phosphates etc., can lead to a
conductivity increase at intermediate mixing ratios. 
Different
theoretical approaches were developed to understand underlying
physical mechanisms causing the  ionic concentration \cite{Anderson/Stuart:1954,
  Maass/etal:1992, Hunt:1994, Maass:1999-p}, the halide doping
\cite{Adams/Swenson:2000} and the mixed glass former effect (MGFE)
\cite{Pradel/etal:2003, Schuch/etal:2009, Schuch/etal:2011}.

The MGFE received particular attention in the last decade
\cite{Schuch/etal:2009, Wang/etal:2018}. Upon mixing of two or more
glass formers, the ionic conductivity can increase (positive MGFE)
\cite{Haynes/etal:2009} or decrease (negative MGFE)
\cite{Martin/etal:2015, Martin/etal:2019}. Many studies in particular
were conducted for alkali borophosphate glasses
\cite{Anantha/Hariharan:2005, Zielniok/etal:2007, Schuch/etal:2011,
  Schuch/etal:2012, Karlsson/etal:2015}.

In a former study, our research group developed a theoretical approach
to explain the MGFE in sodium borophosphate glasses with compositions
0.4Na$_2$O-0.6[$x$B$_2$O$_3$--$(1$-$x)$P$_2$O$_5$] in the whole range
of borate to phosphate mixing $0\le x\le 1$.\cite{Schuch/etal:2011} In
this approach, a thermodynamical model was first introduced to
describe how relative amounts of different types of network forming
units (NFUs) change with $x$. Excellent agreement with experimental
results from magic angle spinning nuclear magnetic resonance (MAS NMR)
was obtained. Based on the NFU concentrations, energy landscapes for
the sodium ion migration were constructed, and the ionic transport
studied by extensive kinetic Monte Carlo (KMC) simulations. These
simulations allowed the extraction of activation energies from Arrhenius
plots of long-time sodium diffusion coefficients. The variation of the
activation energies with $x$ showed good agreement with experimental
results for conductivity activation energies obtained from impedance
spectroscopy. Only one disorder parameter was fitted in the theory,
when assuming a given concentration of vacant ionic sites in the glass
network. This parameter specifies additional fluctuations in the
energetics of individual ion jumps.

Here we revisit the theoretical approach and give explicit expressions
for the NFU concentrations, including additional results in the case
of low alkali concentrations. Thereafter we show that it is possible
to calculate conductivity activation energies analytically from the
landscape construction, which compare well with the KMC results and
experimental data.

We first discuss that evaluating differences between percolation and
Fermi energies, as suggested earlier \cite{Maass/etal:1996,
  Maass:1999-p, Kirchheim:2000}, yields approximate results only. An
accurate theory is provided by an effective mapping to a disordered
conductance network \cite{Ambegaokar/etal:1971, Tyc/Halperin:1989}, or
equivalently, to a single-particle hopping model in a landscape with
spatially varying jump barriers. This means that a purely analytical
approach, with just a numerical determination of Fermi energies and
critical percolation barriers, is able to connect information on the structure
(NFU concentrations) with transport properties (activation
energies) in very good agreement with experiments.  We find it
surprising that this is possible in a complex disordered system such
as the sodium borophosphate glasses.

The analytical treatment allows avoiding extensive computer
simulations and for checking the robustness of the theoretical approach
against variations of parameters within reasonably limits, in particular 
the concentration of vacant ionic sites.  Moreover, one can quickly apply 
the approach now to other borophosphate glasses. We demonstrate this for lithium
borophosphate glasses with compositions
0.33Li$_2$O-0.67[$x$B$_2$O$_3$--$(1$-$x)$P$_2$O$_5$], where
experimental results for NFU concentrations and activation energies
were reported in Ref.~\citenum{Larink/etal:2012}.

\section{Network Former Unit Concentrations}
\label{sec:nfu}
\begin{figure}[b!]
\centering
\includegraphics[width=0.48\textwidth]{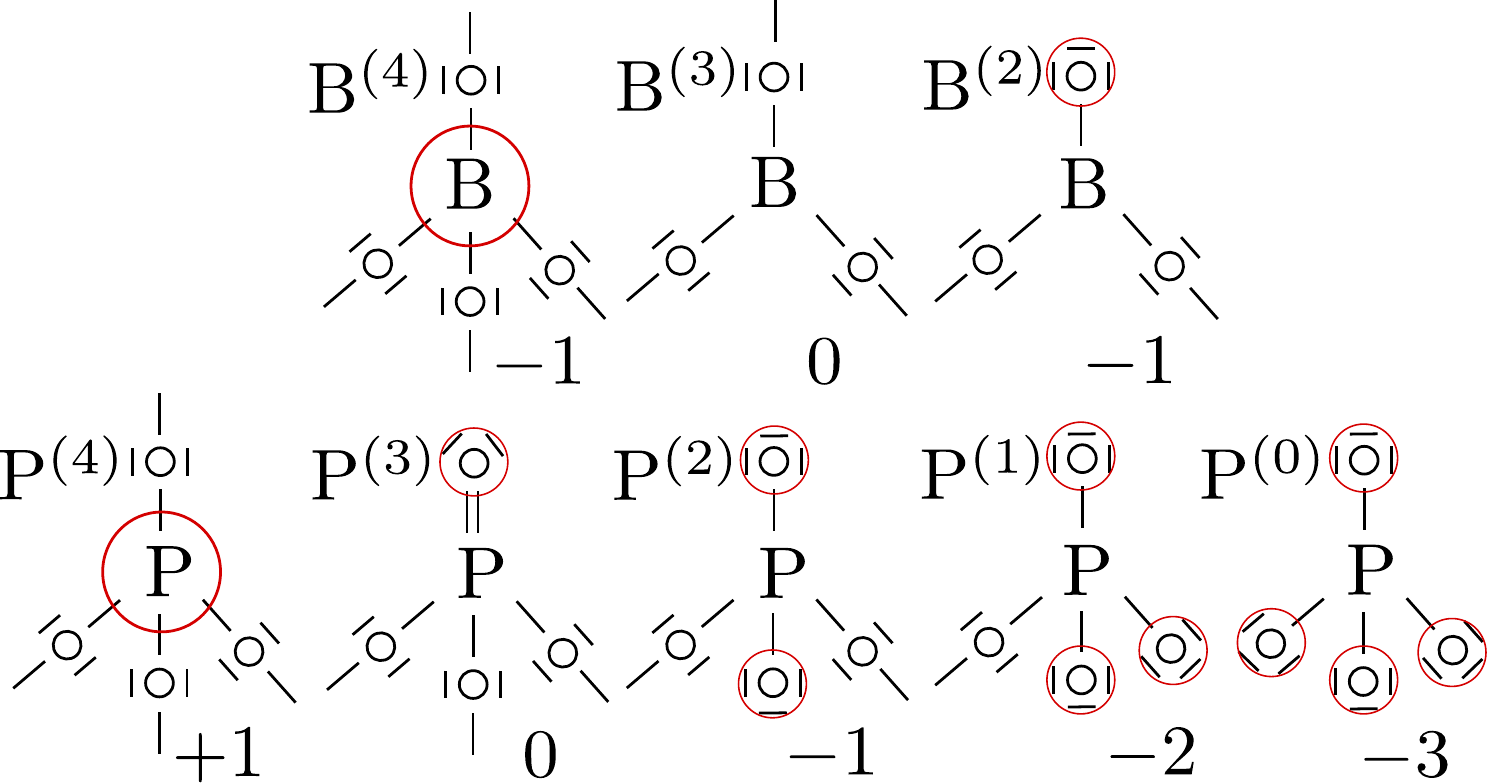}
\caption{Sketches of the different NFUs forming a network in
  borophosphate glasses. The upper and lower rows show the B$^{(n)}$
  and P$^{(n)}$ units with $n$ bridging oxygens, respectively. The
  numbers indicate the charges of the species (in units of the
  elementary charge) and the red circles illustrate how the total
  charge can be viewed as (partially) distributed among nonbridging
  oxygens or among all oxygens in the absence of nonbridging ones.}
\label{fig:nfu}
\end{figure}

MAS-NMR studies for borophosphate glasses \cite{Zielniok/etal:2007,
  Rinke/Eckert:2011} with compositions
$yM_2$O-$(1\!-\!y)$[$x$B$_2$O$_3$-$(1\!-\!x)$P$_2$O$_5$] ($M$: alkali
ion, $y$: molar fraction of alkali oxide) have shown the occurrence of
eight different NFU types, see Fig.~\ref{fig:nfu}: neutral trigonal
B$^{(3)}$ units with three bridging oxygens (bOs), trigonal B$^{(2)}$
units with two bOs and one negatively charged nonbridging oxygen
(nbO), negatively charged tetrahedral B$^{(4)}$ units with four bOs,
and tetrahedral phosphate units P$^{(n)}$, $n=0,\ldots4,$, with $n$
bOs, $(4-n)$ nbOs, and charges $(n-3)$ (in units of the elementary
charge).  The concentration of these different NFU types depends on
the overall glass composition.

%%%%%%%%%%%%%%%%%%%%%%%%%%%%%%%%%%%%%%%
%%%%%%%%%%%%%%%%%%%%%%%%%%%%%%%%%%%%%%%

%%%%%%%%%%%%%%%%%%%%%%%%%%%%%%%%%%%%%%%
\begin{table*}[t!]
\renewcommand{\arraystretch}{1.3}
\begin{tabular}{|c|c|c|c|c|c|}
\hline &&&&& \\[-2.5ex]
NFU & $0\!\le\! x\!\le\! x_1$ & $x_1\!\le\! x\!\le\! x_2^{\rm I}$ & $x_2^{\rm I}\!\le\! x\!\le\! x_2^{\rm II}$  & $x_2^{\rm II}\!\le\! x\!\le\! x_2$ & $x_2\!\le\! x\!\le\! 1$ \\[1ex]
\hline &&&&& \\[-2.5ex]
B$^{(2)}$ & 0 & 0 & 0 & 0 & $\displaystyle \frac{-21\!+\!7\{M\}}{6}\!+\!3x$\\[2ex]
\hline &&&&& \\[-2.5ex]
B$^{(3)}$ & 0 & $\displaystyle -\frac{9\!-\!\{M\}}{20}\!+\!x$ & $\displaystyle \frac{-13\!+\!5\{M\}}{16}\!+\!\frac{5}{4}\,x$ & $\displaystyle -\frac{3-\{M\}}{6}+x$ & $3\!-\!\{M\}\!-\!2x$\\[2ex]
\hline &&&&& \\[-2.5ex]
B$^{(4)}$ & $x$ & $\displaystyle \frac{9\!-\!\{M\}}{20}$ & $\displaystyle \frac{13\!-\!5\{M\}}{16}\!-\!\frac{1}{4}\,x$ & $\displaystyle \frac{3\!-\!\{M\}}{6}$ & $\displaystyle \frac{3\!-\!\{M\}}{6}$\\[2ex]
\hline &&&&& \\[-2.5ex]
P$^{(0)}$ & 0 & 0 & 0 & $\displaystyle\frac{-15\!+\!7\{M\}}{6}\!+\!2x$ & $1\!-\!x$\\[2ex]
\hline &&&&& \\[-2.5ex]
P$^{(1)}$ & 0 & 0 & $\displaystyle \frac{-29\!+\!21\{M\}}{16}\!+\!\frac{5}{4}\,x$ & $\displaystyle \frac{21\!-\!7\{M\}}{6}\!-\!3x$ & 0\\[2ex]
\hline &&&&& \\[-2.5ex]
P$^{(2)}$ & $\{M\}\!-\!x$ & $\displaystyle \frac{-9\!+\!21\{M\}}{20}$ & $\displaystyle \frac{45\!-\!21\{M\}}{16}\!-\!\frac{9}{4}\,x$ & 0 & 0\\[2ex]
\hline &&&&& \\[-2.5ex]
P$^{(3)}$ & $1\!-\!\{M\}$ & $\displaystyle \frac{29\!-\!21\{M\}}{20}\!-\!x$ & 0 & 0 & 0\\[2ex]
\hline
\end{tabular}
\caption{Theoretically predicted fractions of the number of NFUs of
  different types relative to the total number of NFUs in alkali
  borophosphate glasses of compositions
  $yM_2$O-$(1\!-\!y)$[$x$B$_2$O$_3$-$(1\!-\!x)$P$_2$O$_5$] with
  $y\ge3/10$ ($\{M\}=y/(1-y)\ge3/7$).  The NFU fractions exhibit
  different dependencies on the alkali ion fraction $\{M\}=y/(1-y)$ in
  various regimes of the borate to phosphate mixing ratio $x$.  The
  borders of the different regimes are given by $x_1=(9-\{M\})/20$,
  $x_2^{\rm I}=(29-21\{M\})/20$, $x_2^{\rm II}=(45-21\{M\})/36$, and
  $x_2=7/6-7\{M\}/18$.}
\label{tab:nfu-equations1}
\end{table*}
%%%%%%%%%%%%%%%%%%%%%%%%%%%%%%%%%%%%%%%

\begin{table*}[t!]
\renewcommand{\arraystretch}{1.3}
\begin{tabular}{|c|c|c|c|c|}
\hline &&&& \\[-2.5ex]
NFU & $0\!\le\! x\!\le\! x_1$ & $x_1\!\le\! x\!\le\! x_2$ & $x_2\!\le\! x\!\le\! x_3$  & $x_3\!\le\! x\!\le\! 1$  \\[1ex]
\hline &&&& \\[-2.5ex]
B$^{(3)}$ & 0 & 0 & $\displaystyle -\frac{3-\{M\}}{6}+\!x$ & $-\!1\!-\!\{M\}\!+\!2x$ \\[2ex]
\hline &&&& \\[-2.5ex]
B$^{(4)}$ & $x$ & $x$ & $\displaystyle \frac{3\!-\!\{M\}}{6}$ & $1\!+\!\{M\}\!-\!x$ \\[2ex]
\hline &&&& \\[-2.5ex]
P$^{(2)}$ & $\{M\}\!-\!x$ & 0 & 0 & 0 \\[2ex]
\hline &&&& \\[-2.5ex]
P$^{(3)}$ & $1\!-\!\{M\}$ & $1\!+\!\{M\}\!-\!2x$ & $\displaystyle \frac{3\!+\!7\{M\}}{6}\!-\!x$ & 0 \\[2ex]
\hline &&&& \\[-2.5ex]
P$^{(4)}$ & 0 & $-\{M\}\!+\!x$ & $\displaystyle \frac{3\!-\!7\{M\}}{6}\!$ & $1\!-\!x$  \\[2ex]
\hline
\end{tabular}
\caption{Theoretically predicted NFU fractions as in
  Table~\ref{tab:nfu-equations1} for $y\le3/10$
  ($\{M\}=y/(1-y)=\le3/7$).  The borders of the different regimes are
  given by $x_1=\{M\}$, $x_2=(3-\{M\})/6$, and $x_3=(3+7\{M\})/6$, see
  Eqs.~\eqref{eq:x1-border-smallM}, \eqref{eq:x2-border-smallM}, and
  \eqref{eq:x3-border-smallM}.}
\label{tab:nfu-equations2}
\end{table*}

We denote by $\{X\}$ ($X=\mathrm{B}^{(n)}$, $n=2$, 3, 4, or
$\mathrm{P}^{(n)}$, $n=0,\ldots,4$) the fraction of NFUs with respect
to the network forming cations, {\it i.e.}
\begin{equation}
\{X\}=\frac{[X]}{[\mathrm{B}]+[\mathrm{P}]}
\end{equation}
The molar fractions are
$[X]=([B]+[P])\{X\}=(1-y)\{X\}$. 
The total boron content, the
total phosphorus content, and the charge neutrality give the
following constraints:
\begin{subequations}
\begin{align}
&\sum_{n=2}^4 \{{\rm B}^{(n)}\}=x\,,
\label{eq:boron-constraint}\\
&\sum_{n=0}^4 \{{\rm P}^{(n)}\}=(1-x)\,,
\label{eq:phosphorous-constraint}\\
&\{{\rm B}^{(4)}\}+\{{\rm B}^{(2)}\}+\{{\rm P}^{(2)}\}+2\{{\rm
P}^{(1)}\}+
3\{{\rm P}^{(0)}\}\nonumber\\
&\hspace{5em}=\{M\}+\{{\rm P}^{(4)}\}\,,
\label{eq:charge-constraint}
\end{align}
\label{eq:constraints}
\end{subequations}
where 
\begin{equation}
\{M\}=\frac{y}{1-y}\,.
\end{equation}
Equations~\eqref{eq:boron-constraint}-\eqref{eq:charge-constraint} are
three determining equations for the eight unknown NFU fractions.

To obtain a complete set of determining equations, a thermodynamic
model was developed \cite{Schuch/etal:2011}, which is based on
different formation enthalpies $G_X$ of the NFUs $X$.  The differences
between these formation enthalpies can be related to the charge
delocalization in the NFUs, with lower formation enthalpies for higher
delocalization.

%%%%%%%%%%%%%%%%%%%%%%%%%%%%%%%%%%%%%%%
\begin{figure}[t!]
\centering
\includegraphics[width=0.433\textwidth]{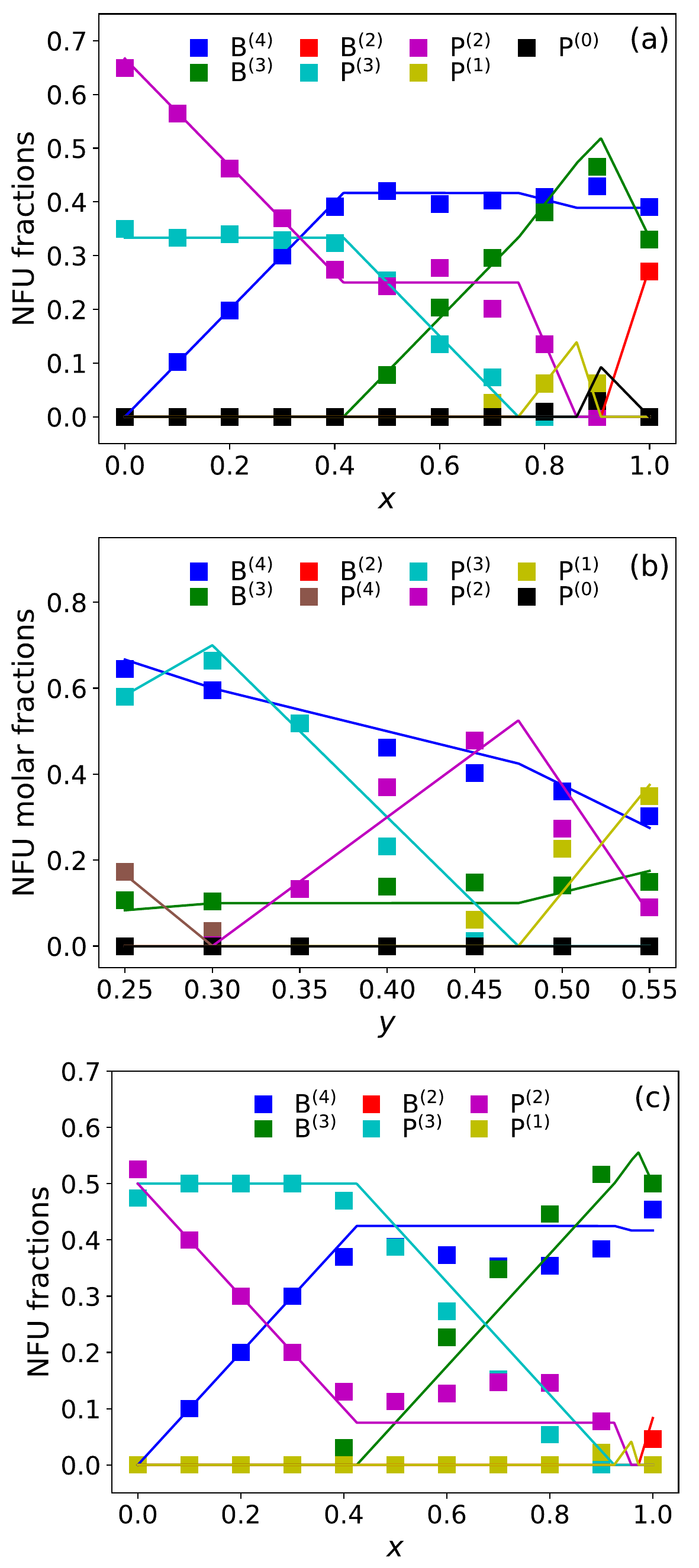}
\caption{Predicted NFU fractions from Table~1 (solid lines) in
  comparison with experimental MAS-NMR data (filled squares) for
  various borophosphate glasses: (a) Glasses with compositions
  0.4Na$_2$O-0.6[$x$B$_2$O$_3$--$(1$-$x)$P$_2$O$_5$] with experimental
  data from Ref.~\citenum{Zielniok/etal:2007} 
  for $0\le x\le 0.9$ and
  from Ref.~\citenum{Michaelis/etal:2007}
  for $x=1$; (b) Glasses with
  compositions $y$Na$_2$O-$2(1$-$y)$[BPO$_4$] with experimental data
  from Ref.~\citenum{Rinke/Eckert:2011}. 
  For $x=0.25$, the theoretical
  values with consideration of P$^{(4)}$ units were used (see text);
  (c) Glasses with compositions
  0.33Li$_2$O-0.67[$x$B$_2$O$_3$--$(1$-$x)$P$_2$O$_5$] with
  experimental data from Ref.~\citenum{Larink/etal:2012}.}
\label{fig:nfu-concentrations}
\end{figure}
%%%%%%%%%%%%%%%%%%%%%%%%%%%%%%%%%%%%%%%

If differences $\Delta G_X$ are significantly larger than the thermal
energy $\kB T_{\rm g}$ at the glass transition ($T_{\rm g}$: glass
transition temperature), a hierarchy results with respect to the
preference of the negatively charged NFUs to compensate for the alkali
ion charges. In the absence of P$^{(4)}$ units, which were not
considered in Ref.~\citenum{Schuch/etal:2011}, the hierarchy corresponds
to an ordering $G_{{\rm B}^{(4)}}<G_{{\rm P}^{(2)}}<G_{{\rm
    P}^{(1)}}<G_{{\rm P}^{(0)}}<G_{{\rm B}^{(2)}}$ of the charged NFU
types.  The hierarchy implies that NFU types with higher formation
enthalpy do not occur unless they must form to satisfy all
constraints. In addition to the constraints given by the equations
\eqref{eq:boron-constraint}-\eqref{eq:charge-constraint}, an important
further one is that of forbidden linkages between ${\rm B}^{(4)}$ units
\cite{Beekenkamp:1965}. This can be attributed to the homogeneously
delocalized charge of the B$^{(4)}$ units, which impedes the formation
of an oxygen bridge between them. For small $\Delta G_X$, which are
less than about 4$\kB T_{\rm g}$, it is possible to refine the
treatment by introducing disproportionation reactions between the
respective NFU types.

In Table~\ref{tab:nfu-equations1} we give the fractions $\{X\}$ of the
various NFU types predicted by the theory \cite{Schuch/etal:2011}
without consideration of disproportionation reactions, and for alkali
ion fractions $\{M\}$ larger than a critical value
$\{M\}_\star=3/7\cong0.43$. In this regime of high alkali content, we
predict no ${\rm P}^{(4)}$ units to occur (see below). Using the equations
in Table~\ref{tab:nfu-equations1}, we find good agreement with MAS-NMR
results for various series of alkali borophosphate glasses with
compositions 0.4Na$_2$O-0.6[$x$B$_2$O$_3$--$(1$-$x)$P$_2$O$_5$],
$y$Na$_2$O-$2(1$-$y)$[BPO$_4$], and
0.33Li$_2$O-0.67[$x$B$_2$O$_3$--$(1$-$x)$P$_2$O$_5$], see
Fig.~\ref{fig:nfu-concentrations}.

If $\{M\}$ is smaller than $\{M\}_\star=3/7$ (or
$y=\{M\}/(1+\{M\})<0.3$), one needs to consider the P$^{(4)}$ units
also. This NFU type has not been considered in
Ref.~\citenum{Schuch/etal:2011}{,} but experimental observations give
strong evidence of its occurrence \cite{Rinke/Eckert:2011,
  Michaelis/etal:2013}. At low alkali content, it becomes
energetically favorable to form more negatively charged ${\rm B}^{(4)}$
units than are needed for compensating all charges of alkali ions. The
additional negative charges of B$^{(4)}$ units are compensated by the
positive charges of P$^{(4)}$ units.

The P$^{(4)}$ units should have small formation enthalpies also due to
their high charge delocalization and their large number of four
bOs. One can imagine mutually linked B$^{(4)}$ and P$^{(4)}$ units to
form small crystalline-type cluster configurations, as sketched in
Fig.~12 of Ref.~\citenum{Rinke/Eckert:2011}. The theoretical modeling in
Ref.~\citenum{Schuch/etal:2011} is thus extended by the following
additional requirement: \textit{ The number of {\rm B$^{(4)}$} units
  is maximal under consideration of positively charged {\rm P$^{(4)}$}
  units and the constraints given by the stoichiometry, charge
  neutrality and forbidden {\rm B$^{(4)}$-B$^{(4)}$} linkages.}  For
the same reasons as for the B$^{(4)}$ units, we expect also that
P$^{(4)}$ units do not link to themselves. On  can check that 
this further constraint can always be fulfilled with the NFU fractions 
calculated below.

The requirement of forbidden B$^{(4)}$-B$^{(4)}$ linkages implies
\begin{align}
&4\{{\rm B}^{(4)}\}\leq 3\{{\rm B}^{(3)}\}+2\{{\rm B}^{(2)}\}+4\{{\rm P}^{(4)}\}\nonumber\\
&\hspace{5em}+3\{{\rm P}^{(3)}\}+2\{{\rm P}^{(2)}\}+\{{\rm P}^{(1)}\}
\label{eq:beekenkamp}
\end{align}
Using this relation, the hierarchy of formation enthalpies and the principle of
maximal possible number of B$^{(4)}$ units, we predict four different
regimes~1 to 4 to occur for $\{M\}\le\{M\}_\star$, 
where certain NFU types are replaced by
certain other types.  Refinements are possible by considering
disproportionation reactions but we are not including them in the
following analysis. The four regimes are as follows:

\textit{Regime~1 {\rm ($0\le x\leq x_1$)}: Replacement of {\rm
    P}$^{(2)}$ by {\rm B}$^{(4)}$ units.} In the phosphate glass
($x=0$), the charges of the alkali ions are compensated by the
P$^{(2)}$ units, $\{{\rm P}^{(2)}\}=\{M\}$ and the rest of the network
is formed by the neutral P$^{(3)}$ units, $\{{\rm
  P}^{(3)}\}=1-\{M\}$. When $x$ is increased, P$^{(2)}$ are first
replaced by B$^{(4)}$ units, because B$^{(4)}$ is the most favorable
NFU type for charge compensation (smallest formation enthalpy of the
negatively charged NFU types). The fraction of P$^{(3)}$ units does
not change. Equations~\eqref{eq:boron-constraint}-\eqref{eq:charge-constraint}
then yield
\begin{subequations}
\label{eq:regime1}
\begin{align}
&\{{\rm B}^{(4)}\}=x\,,\label{eq:regime1-a}\\
&\{{\rm P}^{(2)}\}=\{M\}-x\,,\label{eq:regime1-b}\\
&\{{\rm P}^{(3)}\}=1-\{M\}\label{eq:regime1-c}\,.
\end{align}
\label{eq:nfu-frac-1}
\end{subequations}
The regime terminates when all P$^{(2)}$ units are substituted at 
\begin{equation}
x_1=\{M\}\,.
\label{eq:x1-border-smallM}
\end{equation} 
One can easily check that the constraint in Eq.~\eqref{eq:beekenkamp}
is always fulfilled for the fractions in
Eqs.~\eqref{eq:regime1-a}-\eqref{eq:regime1-c} if
$\{M\}\le\{M\}_\star$.

\vspace{0.5ex}
\textit{Regime~2 {\rm ($x_1\le x\leq x_2$)}: Replacement of {\rm
    P}$^{(3)}$ by {\rm P}$^{(4)}$ and {\rm B}$^{(4)}$ units.}
According to the discussion above, more negatively charged B$^{(4)}$
units can form than are needed for compensating the alkali ion
charges.  This holds true as long as B$^{(4)}$-B$^{(4)}$ linkages can
be avoided. The additional negative charges of B$^{(4)}$ units are
compensated by the positive charges of P$^{(4)}$ units. The network is
now formed by B$^{(4)}$, P$^{(3)}$ and P$^{(4)}$ units and
Eqs.~\eqref{eq:boron-constraint}-\eqref{eq:charge-constraint} yield
\begin{subequations}
\begin{align}
&\{{\rm B}^{(4)}\}=x\,,\\
&\{{\rm P}^{(3)}\}=1+\{M\}-2x\,,\\
&\{{\rm P}^{(4)}\}=x-\{M\}\,.
\end{align}
\label{eq:nfu-frac-2}
\end{subequations}
This regime terminates, if condition \eqref{eq:beekenkamp} becomes violated at 
\begin{equation}
x_2=\frac{1}{2}-\frac{\{M\}}{6}\,,
\label{eq:x2-border-smallM}
\end{equation}
i.e.\ when B$^{(4)}$-B$^{(4)}$ linkages can no longer be
avoided. For $\{M\}=\{M\}_\star=3/7$ in particular, it holds $x_2=x_1$
and the regime~2 disappears. This explains why we expect P$^{(4)}$
units to be absent for $\{M\}\ge\{M\}_\star$.

\vspace{0.5ex}
\textit{Regime~3 {\rm ($x_2\le x\leq x_3$)}: Replacement of {\rm
    P}$^{(3)}$ by {\rm B}$^{(3)}$ units.} As the fraction of B$^{(4)}$
units has saturated, B$^{(3)}$ units are replacing P$^{(3)}$ units
now. The network is formed by B$^{(3)}$, B$^{(4)}$, P$^{(3)}$, and
P$^{(4)}$ units. Equations~\eqref{eq:boron-constraint}-\eqref{eq:charge-constraint}
together with condition \eqref{eq:beekenkamp} (as equation) yield
\begin{subequations}
\begin{align}
&\{{\rm B}^{(3)}\}=-\frac{1}{2}+\frac{1}{6}\{M\}+x\,,\\
&\{{\rm B}^{(4)}\}=\frac{1}{2}-\frac{1}{6}\{M\}\,,\\
&\{{\rm P}^{(3)}\}=\frac{1}{2}+\frac{7}{6}\{M\}-x\,,\\
&\{{\rm P}^{(4)}\}=\frac{1}{2}-\frac{7}{6}\{M\}\,.
\end{align}
\label{eq:nfu-conc-m<m*phase_x2}
\end{subequations}
The regime terminates when all P$^{(3)}$ units are replaced at 
\begin{equation}
x_3=\frac{1}{2}+\frac{7}{6}\{M\}\,.
\label{eq:x3-border-smallM}
\end{equation}

\textit{Regime~4 {\rm ($x_3\le x\leq x_4$)}: Replacement of {\rm
    P}$^{(4)}$ by {\rm B}$^{(3)}$ units.}  Eventually B$^{(3)}$ units
replace P$^{(4)}$ units and the number of B$^{(4)}$ units decreases to
keep the overall charge neutrality.  The network is formed by
B$^{(3)}$, P$^{(4)}$ and B$^{(4)}$ units and
Eqs.~\eqref{eq:boron-constraint}-\eqref{eq:charge-constraint} yield
\begin{subequations}
\begin{align}
&\{{\rm B}^{(3)}\}=-1-\{M\}+2x\,,\\
&\{{\rm B}^{(4)}\}=1+\{M\}-x\,,\\
&\{{\rm P}^{(4)}\}=1-x\,.
\end{align}
\end{subequations}
This behaviour continues until at $x=1$ the network is built
by charged B$^{(4)}$ units with fraction $\{{\rm B}^{(4)}\}=\{M\}$
and neutral B$^{(3)}$ units with fraction $\{{\rm B}^{(3)}\}=1-\{M\}$.

\vspace{0.5ex}
The predicted variations of the fractions of NFU types in the four regimes~1-4
are summarized in Table~\ref{tab:nfu-equations2}.  A few experimental
results for NFU fractions when $\{M\}<\{M\}_\star=3/7$, or $y<0.3$,
are shown in Fig.~\ref{fig:nfu-concentrations}(b). They are in good
agreement with the theoretical predictions.

\section{Site Energy Landscapes for Charge Transport}
\label{sec:sel}
The negatively charged NFUs form counter charges for the mobile alkali
ions. If the relative concentrations of NFU types change, the spatial
distribution of the counter charge becomes modified and hence the
energy landscape for the ion migration. To describe changes of
conductivity activation energies $E_{\rm a}$ with the glass
composition in alkali borophosphate glasses, we base our modeling on
the hypothesis that the modification of site energies, i.e.\ the
energies of the mobile ions at their residence sites in the glass
network, plays the decisive role.

In general, when considering a jump of a mobile ion from a site $i$ to
a vacant neighboring site $j$, the energies $\tilde\epsilon_i$ and
$\tilde\epsilon_j$ of the ion at the initial and target site
matter,\footnote{The notation with the tilde is used, because we
  analyze the charge transport finally in a vacancy picture, where the
  vacancies have site energies $\epsilon_i=-\tilde\epsilon_i$, see
  below.} as well as the saddle point energy to be surmounted. All
these energies are affected by the NFUs in the environment of the
sites $i, j$ and by the Coulomb interaction with the other mobile
ions. To assume that the NFU contribution to the site energies is
decisive, requires the spatial variation of the $\tilde\epsilon_i$
caused by the Coulomb contribution to be weak compared to that caused
by the NFUs. In addition, the saddle point energies should vary weakly
in comparison to the site energies.

As for the Coulomb interaction, molecular dynamics simulations suggest
that the assumption of a weakly fluctuating contribution to the
energetics of ion jumps is better justified in a ``vacancy picture''
\cite{Vogel:2004, Lammert/etal:2009, Lammert/Heuer:2010}.  In this
picture, one considers the vacancies, i.e.\ the vacant ion sites as
charge carriers with a negative charge (analogous to hole conduction
in energy bands of semiconductors). Only a small fraction $f_0$ of the
ion sites is found to be vacant in molecular dynamics simulations,
with $f_0$ typical in the range of 3-10\% \cite{Cormack/etal:2002,
  Lammert/etal:2003, Habasaki/Hiwatari:2004, Mueller/etal:2007}. Small
vacancy concentrations can be expected on thermodynamic reasons,
because ``defect concentrations'' should be small in melt-grown
glasses \cite{Dyre:2003}. It was furthermore shown that $f_0\ll1$ is
required to understand the peculiar behavior seen in the internal
friction behavior of mixed alkali glasses \cite{Peibst/etal:2005,
  Maass/Peibst:2006-p}.

\begin{figure}[t!]
\centering
\includegraphics[width=0.48\textwidth]{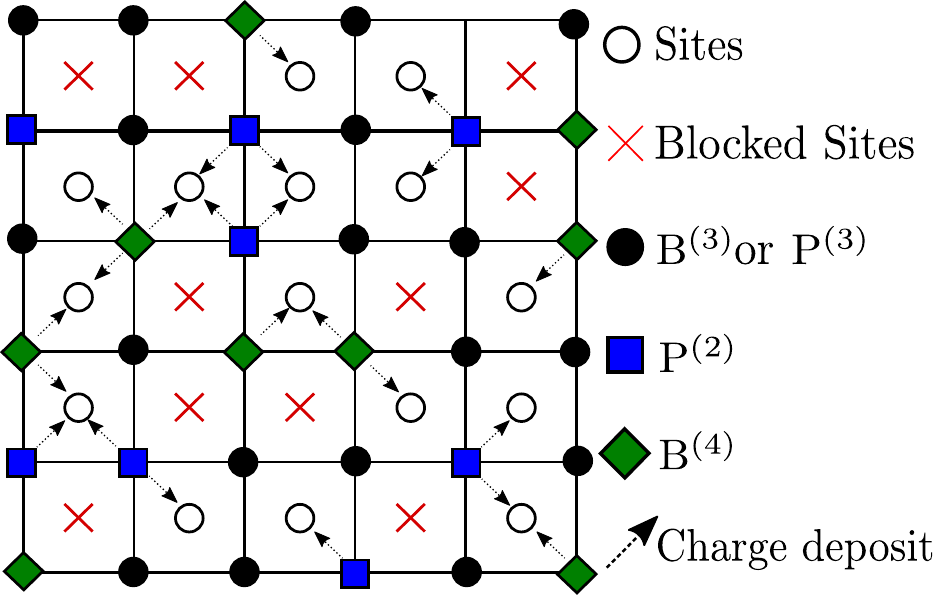}
\caption{Two-dimensional sketch illustrating the M sublattice with
  accessible ion sites (open circles) and blocked sites (crosses), and
  the NFU sublattice with sites randomly occupied by different NFUs
  (with the different species occurring with their respective
  concentrations).  The arrows indicate how the charge of the NFUs
  (see also Fig.~\ref{fig:nfu}) is distributed among the neighboring
  ion sites.}
\label{fig:charge-transfer}
\end{figure}

To construct site energy landscapes in alkali borophosphate glasses,
we follow the approach in Ref.~\citenum{Schuch/etal:2011} and consider
the two simple-cubic sublattices of a body-centered cubic lattice.
 A two-dimensional sketch of these sublattices is shown in Fig.~\ref{fig:charge-transfer}.
One of the sublattices is the ``NFU lattice''.  Its ``NFU sites''
are randomly occupied by NFUs of different types, where each type
occurs with its respective concentration.

On the other sublattice, termed the ``M lattice'', we consider the
ionic motion to take place. A site in this M lattice can be occupied
by at most one mobile ion and ionic motion proceeds by jumps of ions
to vacant nearest neighbor site (as long these are accessible, see
below). As the amount of mobile ions relative to the total number of
NFUs is $\{M\}=y/ (1-y)$ (see Sec.~\ref{sec:nfu}), a fraction
$y/(1-y)$ of all sites in the M lattice is occupied. This requires
$\{M\}=y/(1-y)<1$, meaning that the applicability of the method is
restricted to glass compositions with $y<1/2$. It is possible to
introduce other types of interwoven sublattices to lift this
restriction.  As we consider compositions with $y=0.33$ and $y=0.4$ in
Sec.~\ref{sec:eact} below, corresponding refinements of the approach
are not necessary here.
 
As for the remaining fraction of unoccupied sites in the M lattice, not
all of these sites are considered to be accessible for the mobile ions. This
is because the corresponding fraction of vacant sites would be too
high (51\% for $y=0.33$ and 33\% for $y=0.4$).  In the modeling, we
fix the fraction $f_0$, i.e.\ the number of vacant sites relative to
the number of accessible residence sites.  All remaining unoccupied
sites in the M lattice are blocked.  The fraction of blocked sites
(relative to all sites in the M lattice) is then given by
\begin{align}
f_{\rm bl}=1-\frac{1}{1-f_0}\frac{y}{1-y}\,.
\label{eq:fbl}
\end{align}
To summarize, for a given $f_0$ in the range 3-10\%, we block a
fraction $f_{\rm bl}$ of sites in the M lattice according to
Eq.~\eqref{eq:fbl}. The other sites are the accessible residence
sites, which we refer to as ``ion sites'' in the
following.\footnote{It is possible that finite clusters of ion sites
  arise due to the blocking, which are not connected to the
  percolating cluster of ion sites.}

%%%%%%%%%%%%%%%%%%%%%%%%%%%%%%%%%%%%%%%
%%%%%%%%%%%%%%%%%%%%%%%%%%%%%%%%%%%%%%%

To the ion sites are assigned partial charges in dependence of the
surrounding neighboring NFUs in the NFU lattice, where we take into
account how charge is localized in the different NFU types (see the
red circles in Fig.~\ref{fig:nfu}).  The procedure is illustrated by
the arrows in Fig.~\ref{fig:charge-transfer}.

Consider an NFU of type $\alpha$ with charge $q_\alpha$ and
$k_\alpha$ nbOs that occupies an NFU site, which has $z$ ion sites in
its immediate surroundings. This means that $z$ of its 8 nearest
neighbor sites in the M lattice are accessible for the mobile
ions. Among these $z$ ion sites, we select $k_\alpha$ randomly
and add a partial charge $q_\alpha/k_\alpha$ to them (in the rare case
$z<k_\alpha$, the NFU is interchanged with a charge-neutral one at
another site; all charges of the NFUs can then be distributed as
described).  For a negatively charged ${\rm B}^{(4)}$ unit without nbO,
where the charge $q_{{\rm B}^{(4)}}=-1$ can be viewed as delocalized over
all four oxygens, we add a charge $(-1/z)$ uniformly to all
surrounding ion sites. A ${\rm P}^{(4)}$ unit with $q_{{\rm P}^{(4)}}=+1$ would be
treated analogously, but for $y=0.33$ and $y=0.4$, we have $\{M\}>3/7$
and ${\rm P}^{(4)}$ units do not appear (see Sec.~\ref{sec:nfu}).

All partial charges at an ion site $i$ are summed up to the total charge
\begin{equation}
Q_i=\sum_{j(i)} q_j\,.
\end{equation}
where the sum over $j=j(i)$ runs over all eight NFU sites $j$
surrounding the ion site $i$, and $q_j$ is the partial charge
contribution of the NFU at site $j$ ($q_j=0$ if the NFU at site $j$ is
charge-neutral).  

The total charge $Q_i$ gives rise to a site energy
\begin{equation}
\tilde\epsilon_i^{\,(0)}=V_0\, Q_i\,, 
\label{eq:eps0i} 
\end{equation}
where $V_0$ is an energy scale. This scale can be roughly estimated by
the Coulomb interaction between an alkali ion with positive charge and
an oxygen with negative charge at a distance of about 2\AA. The
precise value $V_0$, however, is not relevant, when we consider
relative changes of activation energies with the glass composition
in Sec.~\ref{sec:eact}.

To account for additional structural and energetic disorder, we add
fluctuations $V_0\eta_i$ to the sites energies, where the $\eta_i$ are
uncorrelated Gaussian random numbers with zero mean and standard
deviation $\sigma_\epsilon$,
\begin{equation}
\tilde\epsilon_i=\tilde\epsilon_i^{\,(0)}+V_0\,\eta_i=V_0\, (Q_i+\eta_i)\,.
\label{eq:epsi}
\end{equation}
The fluctuations have several sources. In particular they should take
into consideration that sites in a glass are not located on a simple
cubic lattice, and that there is an additional contribution from the
Coulomb interaction between mobile ions (see the discussion
above). Moreover, for ion jump rates, whose variation depends only on
the site energies, we can think of the fluctuations to take into
account also spatial variations of the saddle points energies.  

As the structure of the glass changes with the borate to phosphate mixing ratio,
the standard deviation or ``disorder parameter'' $\sigma_\epsilon$ should depend on $x$ also. 
However, according to our hypothesis, the redistribution of counter-charges and associated
modifications of the site energies should be the main cause of 
changes in long-range ion transport properties. Hence, $\sigma_\epsilon$ should vary only weakly with $x$. More precisely, 
its variation with $x$ should be small compared to the variations of the $\tilde\epsilon_i^{(0)}$ in Eq.~\eqref{eq:eps0i}.
In order to see whether the dependence of conductivity activation energies $E_{\rm a}$ on $x$  
can be captured without adjusting $\sigma_\epsilon$ for each $x$, we deliberately constrain 
$\sigma_\epsilon$ to a fixed value. It is the
only fit parameter then, when we calculate normalized values $E_{\rm a}(x)/E_{\rm a}(0)$ for different 
mixing ratios $x$ in the next section.

\section{Conductivity Activation\\ Energies}
\label{sec:eact}
\begin{figure}[t!]
\centering
\includegraphics[width=0.45\textwidth]{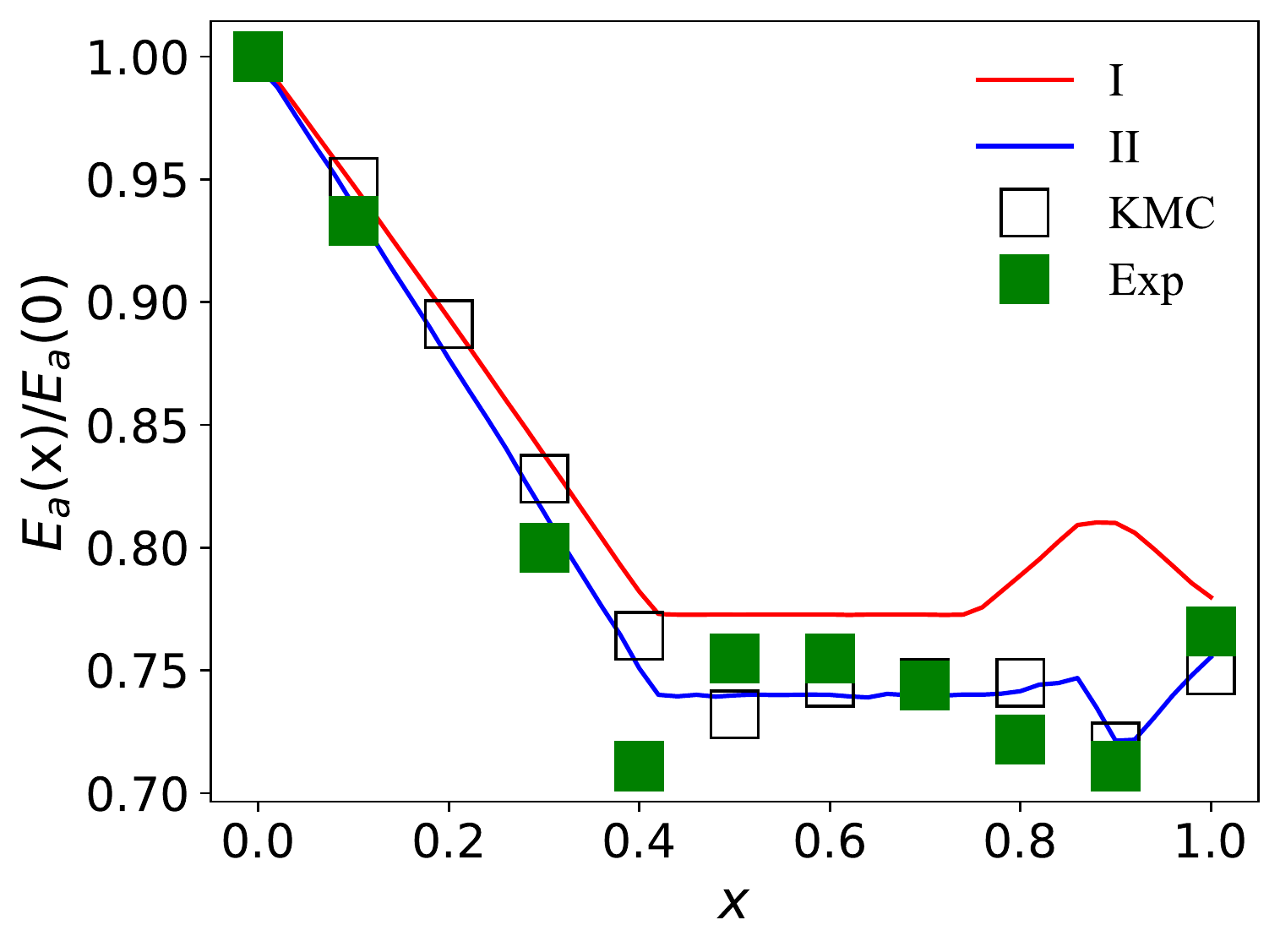}
\caption{Normalized activation energy $E_{\rm a}(x)/E_{\rm a}(0)$ as a
  function of the network former mixing ratio $x$ in sodium
  borophosphate glasses with compositions
  0.4Na$_2$O-0.6[$x$B$_2$O$_3$--$(1$-$x)$P$_2$O$_5$].  The filled
  symbols are experimental data taken from
  Ref.~\citenum{Zielniok/etal:2007} and the open symbols results from KMC
  simulations reported in Ref.~\citenum{Schuch/etal:2011}, the value at
  $x=1$ obtained from the KMC simulations has been recalculated. The
  lines correspond to the analytically calculated activation energies
  from approaches I (red line) and II (blue line).}
\label{fig:activation-energy-sodium}
\end{figure}

In a former study \cite{Schuch/etal:2011}, it was shown that the
change of $E_{\rm a}(x)/E_{\rm a}(0)$
with $x$ in the glasses
0.4Na$_2$O-0.6[$x$B$_2$O$_3$--$(1$-$x)$P$_2$O$_5$] could be
successfully modeled based on the energy landscape construction
presented in Sec.~\ref{sec:sel}, despite of its simplicity. The
activation energies in the respective study were obtained by extensive
kinetic Monte-Carlo (KMC-) simulations with a Metropolis form of the
jump rates of the mobile ions for $f_0=0.1$. Using a disorder
parameter $\sigma_\epsilon=0.3$, good agreement with measured
values was obtained, see the filled (experiment) and open (KMC simulations)
symbols in Fig.~\ref{fig:activation-energy-sodium}. The energy scale $V_0$ 
in real units can be estimated by 
requiring the measured value $E_{\rm a}(0)$ to agree with the simulated
one. This yields $V_0\simeq1.29\,$eV.

Here we will present two analytical approaches I and II to determine
the activation energy and compare the results to those obtained
earlier by the KMC simulations. Approach II 
will be applied also to lithium borophosphate glasses
with compositions
0.33Li$_2$O-0.67[$x$B$_2$O$_3$--$(1$-$x)$P$_2$O$_5$].  For both
analytical approaches we use the vacancy picture, for which the site
energies $\epsilon_i$ have opposite sign,
$\epsilon_i=-\tilde\epsilon_i$.

%%%%%%%%%%%%%%%%%%%%%%%%%%%%%%%%%%%%%%%
%%%%%%%%%%%%%%%%%%%%%%%%%%%%%%%%%%%%%%%

\begin{figure}[t!]
\centering
\includegraphics[width=0.45\textwidth]{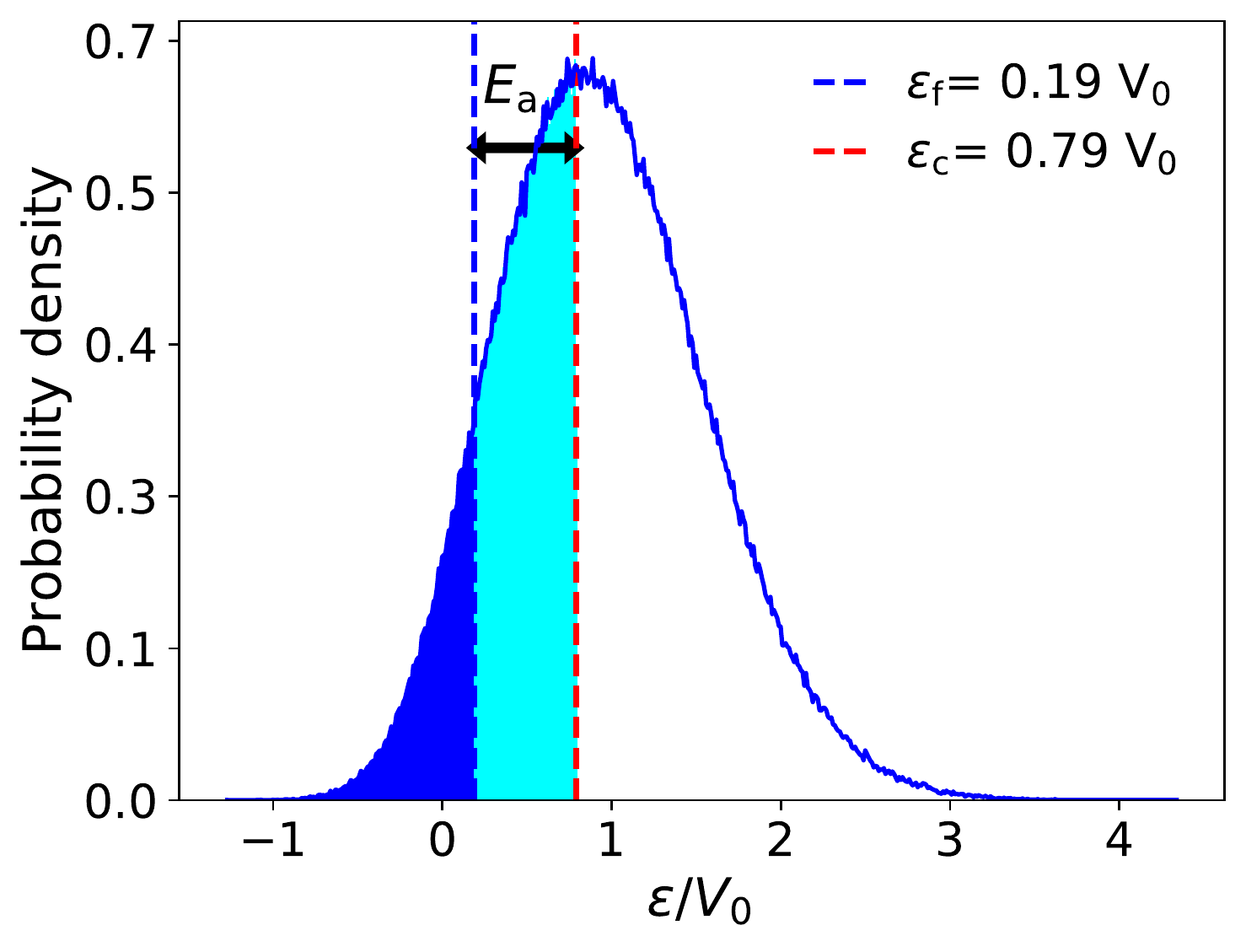}
\caption{Distribution of site energies for vacancies according to the
  energy landscape construction described in Sec.~\ref{sec:sel} for
  the glass 0.4Na$_2$O-0.6[$x$B$_2$O$_3$--$(1$-$x)$P$_2$O$_5$] with
  $x=0.25$. The fraction of vacancies relative to all accessible sites
  is $f_0=0.1$.  The Fermi energy $\epsilon_{\rm f}$ indicates the
  highest energy of vacant sites in thermal equilibrium in the low
  temperature limit $T\to0$.  The critical energy $\epsilon_{\rm c}$
  is the minimal site energy required for all sites with
  $\epsilon_i\le\epsilon_{\rm c}$ to form a percolating path.  The
  difference $(\epsilon_{\rm c}-\epsilon_{\rm f})$ gives the
  activation energy $E_{\rm a}$ in model I.}
\label{fig:example-site-energy-distribution}
\end{figure}

Method I is based on a simple picture that has been used in former
works \cite{Porto/etal:2000, Kirchheim:2000}. It is illlustrated in
Fig.~\ref{fig:example-site-energy-distribution}, where we show the
distribution of site energies resulting from our landscape
construction for the glass
0.4Na$_2$O-0.6[$x$B$_2$O$_3$--$(1$-$x)$P$_2$O$_5$] with $x=0.25$.  In
the low-temperature limit $T\to0$, the vacancies fill up all sites
with energies $\epsilon_i<\epsilon_{\rm f}$, where $\epsilon_{\rm f}$
is the Fermi energy.\footnote{Note that a site can be occupied by at
  most one vacancy, i.e.\ the Fermi statistics applies to the
  vacancies as well.} Due to thermal fluctuations, vacancies can move
through the ``whole system'' (long range diffusion corresponding to
dc-conductivity), where at low temperatures the respective diffusion
path should consist of all sites with energies $\epsilon_i$ below a
critical value $\epsilon_{\rm c}$. This critical value is the lowest
one, where a connected path of sites with energies below a given
threshold $\epsilon_{\rm th}$ can form, i.e.\ $\epsilon_{\rm
  c}=\min_{\epsilon_{\rm th}}(\mbox{set of sites with
  $\epsilon_i\le\epsilon_{\rm th}$ is percolating)}$. The conductivity
activation energy is then given by
\begin{equation}
E_{\rm a}=\epsilon_{\rm c}-\epsilon_{\rm f}\,.
\end{equation}

%%%%%%%%%%%%%%%%%%%%%%%%%%%%%%%%%%%%%%%
%%%%%%%%%%%%%%%%%%%%%%%%%%%%%%%%%%%%%%%

We have determined both the Fermi energy $\epsilon_{\rm f}$ and the
critical energy $\epsilon_{\rm c}$ by generating site energy
landscapes in systems (M lattices) with 100$^3$ sites. The
determination of $\epsilon_{\rm f}$ is straightforward and for
determining the critical energy $\epsilon_{\rm c}$ we used the
Hoshen-Kopelman algorithm of percolation theory
\cite{Hoshen/Kopelman:1976}. Once the algorithm is implemented, the
calculation of one value of $E_{\rm a}$ takes just a few minutes on a
standard personal computer.

The normalized $E_{\rm a}(x)/E_{\rm a}(0)$ calculated for the glasses
0.4Na$_2$O-0.6[$x$B$_2$O$_3$--$(1$-$x)$P$_2$O$_5$] with approach I 
is shown in Fig.~\ref{fig:activation-energy-sodium} (red line) for the fixed
value $f_0=0.1$ formerly used in the KMC simulations.  The agreement
with the KMC results (open symbols) is good for $x\lesssim 0.3$, but
for larger mixing ratios the activation energy is overestimated. At
$x\cong0.9$, the predicted $E_a(x)$ exhibits a local maximum, while
the KMC results indicate a local minimum. Hence, model I does not give
a satisfactory prediction of $E_{\rm a}(x)$ for all $x$.

In approach II, we make use of the fact that the hopping transport
in a Fermionic lattice gas with site energy disorder can be mapped
onto an effective single particle hopping model in a landscape, where
all site energies are equal, and the energetic disorder is completely
transferred to the jump barriers (this model corresponds also to a
random conductance network).\cite{Ambegaokar/etal:1971} The barrier
$\Delta_{ij}$ for a jump from site $i$ to a nearest-neighbor site $j$
is given by
\begin{equation}
\Delta_{ij}=\frac{1}{2}\left(|\epsilon_i-\epsilon_j|+|\epsilon_i-\epsilon_{\rm f}|+|\epsilon_j-\epsilon_{\rm f}|\right)\,.
\label{eq:eps-delta}
\end{equation}
Note that the barriers are symmetric, $\Delta_{ij}=\Delta_{ji}$, and
that the vacancy concentration enters via the Fermi energy
$\epsilon_{\rm f}$.  Bonds in the M lattice connected to blocked sites
are assigned infinite barriers.

%%%%%%%%%%%%%%%%%%%%%%%%%%%%%%%%%%%%%%%
\begin{figure}[t!]
\centering
\includegraphics[width=0.45\textwidth]{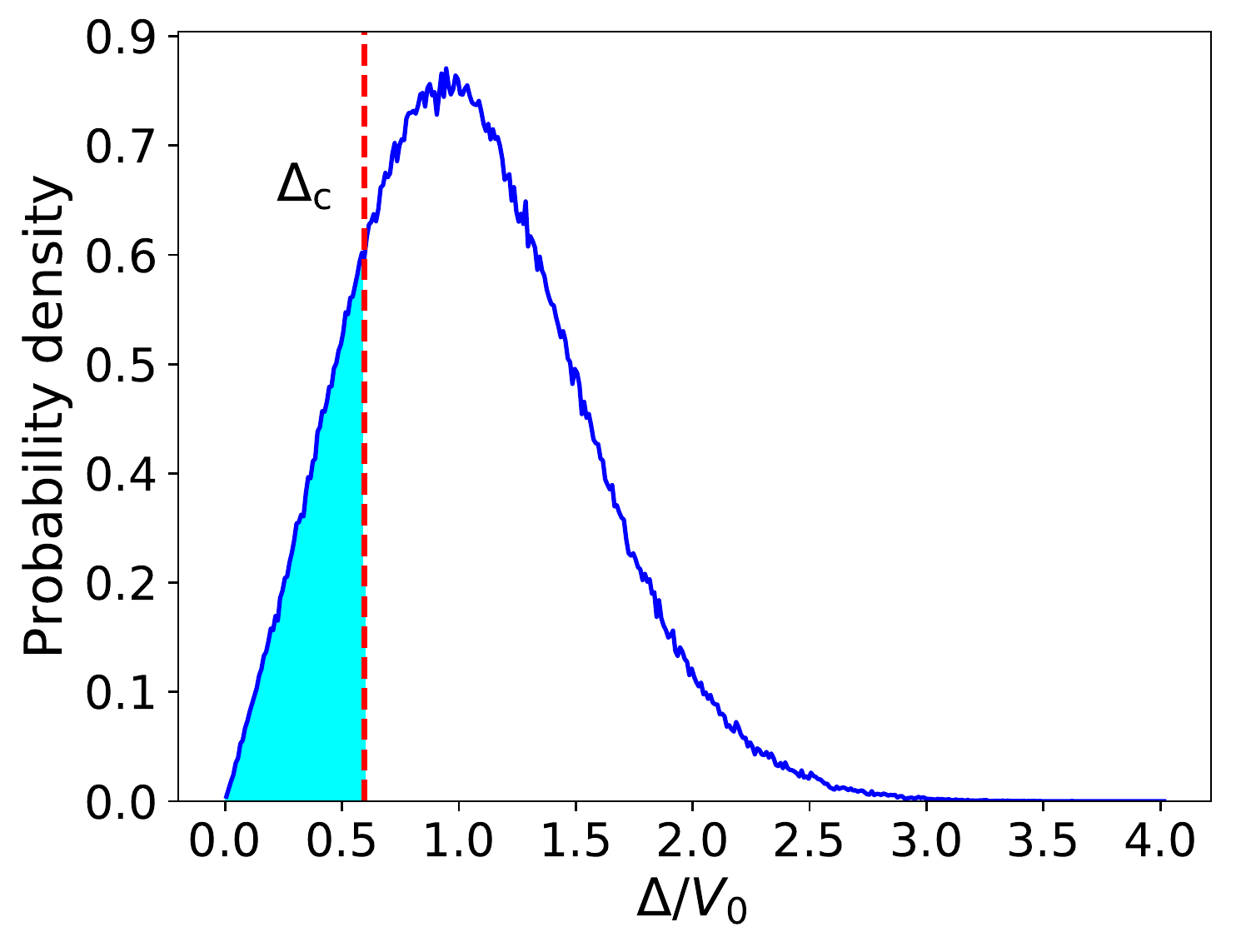}
\caption{Distribution of energy barriers calculated from the site
  energies according to Eq.~\eqref{eq:eps-delta} for the same energy
  landscape as considered in
  Fig.~\ref{fig:example-site-energy-distribution}. The critical
  barrier $\Delta_{\rm c}$ is the minimal barrier energy required for
  all bonds with $\Delta_{ij}\le\Delta_{\rm c}$ to form a percolating
  path. It equals the activation energy $E_{\rm a}$.}
\label{fig:example-barrier-distribution}
\end{figure}
%%%%%%%%%%%%%%%%%%%%%%%%%%%%%%%%%%%%%%%

The activation energy follows by considering all bonds $(ij)$ with
$\Delta_{ij}$ below a threshold value $\Delta_{\rm th}$. It holds
\begin{equation}
E_{\rm a}=\Delta_{\rm c}\,,
\label{eq:eaII}
\end{equation}
where $\Delta_{\rm c}$ is the critical smallest value of the thresholds, where the bonds with $\Delta_{ij}\le\Delta_{\rm th}$
still form a percolating path, i.e.\ $\Delta_{\rm c}=\min_{\Delta_{\rm th}}$(set of bonds $(ij)$ with $\Delta_{ij}\le\Delta_{\rm th}$ is percolating). 

Figure~\ref{fig:example-barrier-distribution} shows as an example the
distribution of barriers calculated from Eq.~\eqref{eq:eps-delta} for
the same energy landscape used in
Figure~\ref{fig:example-site-energy-distribution}. The critical
barrier $\Delta_{\rm c}$ indicated by the red vertical line was
determined again by using the Hoshen-Kopelman algorithm, where this
time the bond percolation was evaluated. 

The activation energies from approach II  are shown
in Fig.~\ref{fig:activation-energy-sodium} as blue line and agree well
with the KMC results for all mixing ratios $x$. Small deviations can
be attributed to uncertainties in the $E_{\rm a}$ values calculated
from the KMC simulations. Hence, conductivity activation energies can now
be calculated very quickly
without the need to perform extensive KMC simulations for various
temperatures and subsequently extracting $E_{\rm a}$ from Arrhenius
plots.

%%%%%%%%%%%%%%%%%%%%%%%%%%%%%%%%%%%%%%%
\begin{figure}[t!]
\centering
\includegraphics[width=0.45\textwidth]{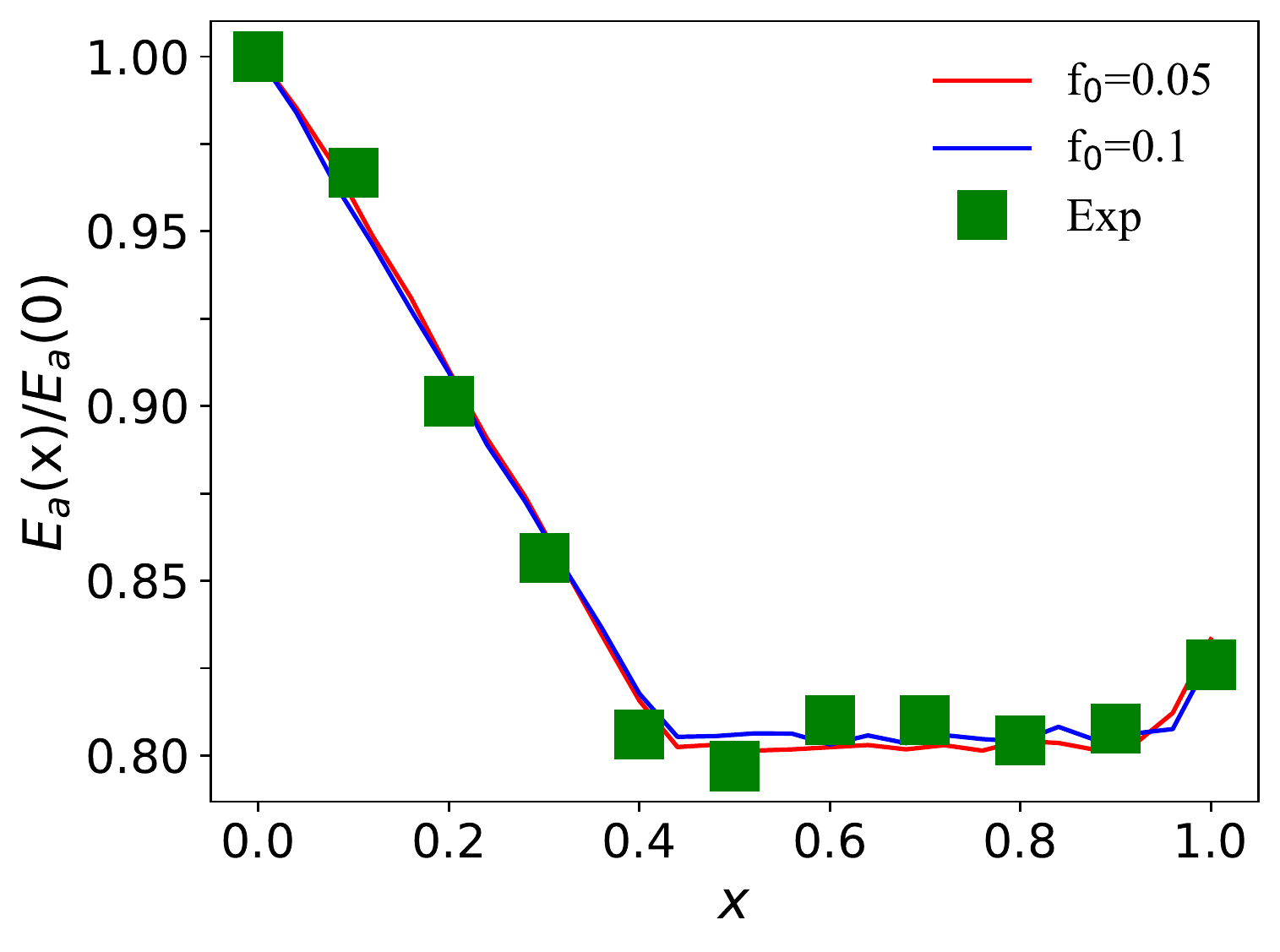}
\caption{Normalized activation energy $E_{\rm a}(x)/E_{\rm a}(0)$ as a
  function of the network former mixing ratio $x$ in lithium
  borophosphate glasses with compositions
  0.33Li$_2$O-0.67[$x$B$_2$O$_3$--$(1$-$x)$P$_2$O$_5$].  The filled
  symbols are experimental data taken from
  Ref.~\citenum{Larink/etal:2012} and the lines analytically calculated
  activation energies with approach II based on the energy landscape
  construction described in Sec.~\ref{sec:sel} for $f_0=0.05$ (red
  line) and $f_0=0.1$ (blue line). The disorder parameter (standard
  deviation) $\sigma_\epsilon$ for the Gaussian fluctuations in
  Eq.~\eqref{eq:epsi} was chosen to match the experimental value
  $E_{\rm a}(1)/E_{\rm a}(0)$ ($\sigma_\epsilon=0.44$ for
  $f_0=0.05$ and $\sigma_\epsilon=0.49$ for $f_0=0.1$).}
\label{fig:activation-energy-lithium}
\end{figure}
%%%%%%%%%%%%%%%%%%%%%%%%%%%%%%%%%%%%%%%

Let us apply the procedure to another series of alkali borophosphate
glasses, namely the lithium conducting ones with compositions
0.33Li$_2$O-0.67[$x$B$_2$O$_3$--$(1$-$x)$P$_2$O$_5$]. Experimental results
 for NFU
concentrations (see Fig.~\ref{fig:nfu-concentrations}) and
conductivity activation energies for these glasses were reported in
Ref.~\citenum{Larink/etal:2012}. To construct site energy landscapes for them, 
we take two fixed values $f_0=0.05$ and $f_0=0.1$ of
the vacancy fraction in order to see, whether the specific value of
$f_0$ is important.  To determine the disorder parameter
$\sigma_\epsilon$, we require $E_{\rm a}(1)/E_{\rm a}(0)$ to match the
experimental value for each of the two considered $f_0$ values. This
means that $\sigma_\epsilon$ is solely determined by the
phosphate ($x=0$) and borate ($x=1$) glass.  The variation of $E_{\rm
  a}(x)/E_{\rm a}(0)$ with the network former mixing ratio is then a
prediction of the theoretical modeling, i.e.\ not affected by any
parameter fitting.

We find $\sigma_\epsilon=0.44$ for $f_0=0.05$ and a slightly larger
value $\sigma_\epsilon=0.49$ for $f_0=0.1$.  The predicted behavior
for $E_{\rm a}(x)/E_{\rm a}(0)$ as a function of $x$ (red line for
$f_0=0.05$, blue line for $f_0=0.1$) is shown in
Fig.~\ref{fig:activation-energy-lithium} and compares very well with
the measured data (symbols). 
When calculating $V_0$ in real units by requiring the measured value $E_{\rm a}(0)$ 
to agree with the simulated one, we obtain the estimates $V_0\simeq0.69\,$eV for 
$f_0=0.05$ and $V_0\simeq0.80\,$eV for $f_0=0.1$.
Let us note that we checked 
also the robustness against changes of $f_0$ in our modeling of the sodium
borophosphate glasses considered in Fig.~\ref{fig:activation-energy-sodium}.

\section{Summary and Conclusions}
\label{sec:summary}
For alkali borophosphate glasses, we derived analytical results for
NFU concentrations, constructed site energy landscapes based on the
concentrations and charges of the NFUs, and calculated conductivity
activation energies from these landscapes by applying a method, which
requires only a simple and quick determination of Fermi and critical
percolation energies. The results for the predicted NFU concentrations
and activation energies compare well with experimental data. We thus
regard our theoretical approach as a promising step forward toward
relating structure to transport properties in ionically conducting glasses,
and toward developing a theoretical understanding of ion transport in
glasses with quantitive predictive power.

There is plenty of room for further developments of the approach.  One
important refinement will be to incorporate spatial correlations
between the NFUs of the various types in the energy landscape
construction. These spatial correlations can be measured, for example,
by applying dipolar recoupling methods in rotational-echo
double-resonance (REDOR) measurements of advanced solid state NMR
\cite{Zhang/Eckert:2006, Rinke/Eckert:2011, Larink/etal:2012,
  Eckert:2018} (for a recent review of solid state NMR studies on
borophophate glasses, see Ref.~\citenum{Tricot/etal:2020}). Another point
is to evaluate aspects of the constructed energy landscapes more
directly with respect to experimental observations, as, for example,
those obtained from the recently developed method of charge attachment
induced transport (CAIT) \cite{Schaefer/Weitzel:2018,
  Schaefer/etal:2019}. Further improvements concern a more detailed
modeling of the local jump dynamics with explicit consideration of
saddle point energies and disorder in the spatial arrangements of ion
sites.

The mapping of the (many-body) jump model with varying site energies
to a single particle model with varying jump barriers moreover
provides insights, why the random barrier model \cite{Dyre:1988,
  Dyre/Schroeder:2000} is successful to capture the quasi-universal
scaling behavior of conductivity spectra in glasses with one type of
mobile ion \cite{Roling/etal:1997, Sidebottom:1999, Ghosh/Pan:2000}. A
corresponding mapping is not possible if more than one type of mobile
ion is present, as, e.g., in mixed alkali glasses. One may thus expect
that conductivity scaling in such glasses is not observed, which is
indeed the case \cite{Cramer/etal:2002}. Calculations of conductivity
spectra and preexponential factors of dc conductivities will be
presented in upcoming work.

\begin{acknowledgement}
%Please use ``The authors thank \ldots'' rather than ``The
%authors would like to thank \ldots''.
P.M.\ thanks H.\ Eckert and S.~W.\ Martin
for very valuable discussions. This work has been funded by the Deutsche Forschungsgemeinschaft (DFG, Project No. 428906592). We sincerely thank the members of the DFG Research Unit FOR 5065 for fruitful discussions.
\end{acknowledgement}

%\bibliography{ionicglasses}

\providecommand{\latin}[1]{#1}
\makeatletter
\providecommand{\doi}
  {\begingroup\let\do\@makeother\dospecials
  \catcode`\{=1 \catcode`\}=2 \doi@aux}
\providecommand{\doi@aux}[1]{\endgroup\texttt{#1}}
\makeatother
\providecommand*\mcitethebibliography{\thebibliography}
\csname @ifundefined\endcsname{endmcitethebibliography}
  {\let\endmcitethebibliography\endthebibliography}{}

\end{document}